\documentclass[prd,longbibliography,nofootinbib,preprintnumbers,twocolumn,showpacs,floatfix,superscriptaddress]{revtex4-1}

\usepackage[utf8]{inputenc}      
\usepackage{amsmath, amsthm, amssymb}
\usepackage{graphicx}
\usepackage{color}
\usepackage{hyperref}

\newtheorem*{prop}{Proposition}

\begin{document}

\preprint{CPPC-2020-04}

\title{Renormalization group  flows and emergent symmetries}

\author{Zurab Berezhiani}
 \email{zurab.berezhiani@aquila.infn.it}
 \affiliation{
Dipartimento di Scienze Fisiche e Chimiche, Universit\`{a} di L'Aquila, 67100 Coppito, L'Aquila, Italy }
\affiliation{INFN, Laboratori Nazionali del Gran Sasso, 67010 Assergi, L'Aquila, Italy }
\author{Maicol Di Giambattista}
 \affiliation{
Dipartimento di Scienze Fisiche e Chimiche, Universit\`{a} di L'Aquila, 67100 Coppito, L'Aquila, Italy }
\affiliation{INFN, Laboratori Nazionali del Gran Sasso, 67010 Assergi, L'Aquila, Italy }
\author{Alessio Maiezza}
 \affiliation{
Dipartimento di Scienze Fisiche e Chimiche, Universit\`{a} di L'Aquila, 67100 Coppito, L'Aquila, Italy }
\author{Archil Kobakhidze}
\affiliation{Sydney Consortium for Particle Physics and Cosmology, \\
 School of Physics, The University of Sydney, NSW 2006, Australia
}

\begin{abstract}
We discuss the following proposition: Renormalization Group flow of quantum theory with a biased symmetry exhibits a fixed hypersurface at which the symmetry is exact. Such emergent symmetries may have important phenomenological implications, including supersymmetric models, gauge theories, and massive gravity.
Most interesting example is an emergent supersymmetry in non-abelian gauge theories
with appropriate field content,
in the IR limit  i.e. strong coupling regime.
\end{abstract}

\maketitle

\section{Emergent symmetries}
Ever since the seminal works by Noether \cite{Noether:1918zz}, Weyl \cite{Weyl:1927vd},
Heisenberg \cite{Werner} and Wigner \cite{Wigner:1939cj}, the role of symmetries in particle physics
is ubiquitous. Symmetries provide the classification of particles,
dictate conservation laws in their interactions, are instrumental in solving dynamical problems, etc.
Some known symmetries have fundamental character, e.g. gauge symmetries of the Standard Model (SM).
These are the precise symmetries of the Lagrangian
used as postulated theoretical inputs when building the particle models.
These symmetries can be broken spontaneously (by the vacuum state) but not explicitly.
One also often deals with the controllable breaking of symmetries, such as spontaneous
and explicit soft or anomalous breaking of symmetries.
Some of the global symmetries emerge accidentally owing to the theoretical structures  dictated by postulated fundamental symmetries
(as e.g. baryon and lepton symmetries or isospin symmetry in the SM context)  and in principle are {\it approximate} symmetries.
Some other symmetries as conformal symmetry can be exact symmetries of the classical Lagrangian but are broken by the renormalization group flow of the dimensionless constants since the mass scale emerges due to dimensional transmutation.

Here we would like to discuss another, in our opinion important class of emergent symmetries.
These are symmetries that manifest themselves only at some scales in {\it a priori} asymmetric theory.
The central result of this paper is the following

\begin{prop}
Consider a quantum field theory with a set of fields $\Phi$  and parameters (coupling constants)
$\lambda_{k}$ which is described by an action $S(\Phi, \lambda_{k},\mu)$
at a renormalization scale $\mu$.
Under the renormalization group (RG) evolution, the theory flows towards a fixed hypersurface
given by $f_i(\lambda_k)=0$  in the parameter space at which $\beta$-function of the constraint $f_i$
 vanishes and the theory exhibits an enhanced symmetry.
\end{prop}

The proof is rather straightforward. The enhanced symmetry under the constraints  $f_i(\lambda_k)=0$ on the theory parameters implies that the variation of the generating functional of the constrained theory  $Z=\int D\Phi\,\delta[f_i(\lambda_k)]\,{\rm exp}[iS]$ vanishes under the symmetry transformations $\delta Z=0$, i.e.
\begin{equation}
\delta S + \mathcal{A}=i\int d^dx\,c_i\,f_i(\lambda_k)~,
\end{equation}
where $\mathcal{A}$ is given by an anomalous variation of the functional measure,
$\mathcal{A}=\ln \delta(D\Phi)$ and $c_i$ are the auxiliary Lagrange  multipliers
that implement the constraints $f_i=0$.
This variation does not depend on the renormalization scale, that is
\begin{equation}
\frac{dc_i}{dt}\,f_i+c_i\,\frac{df_i}{dt }=0~,
\end{equation}
where $t=\ln\mu$. The last equation in turn implies
\begin{equation}
 \frac{df_i}{dt }=\beta_{f_i}/ (4\pi)^2 = 0
\end{equation}
on the constraint hypersurface $f_i=0$.
Therefore, the constraint are fixed hypersurfaces of the RG equations.

This observation has significant implications for our understanding of the role of biased symmetries. In effect, any {\it a priori} asymmetric theory exhibits an emergent symmetry, providing the symmetry-enhanced hyper-surface exists. The emergent symmetries are a common feature of many condensed matter systems, while remain less explored in high energy physics. We believe they can provide new insights into fundamental problems. One such is the naturalness of physical theories, which according to common lore demands some enhanced symmetries \cite{tHooft:1979rat}. These symmetries may be emergent rather than the fundamental feature of the theory. The emergent nature of spacetime symmetries such as relativistic invariance \cite{Chadha:1982qq} and/or supersymmetry \cite{Iliopoulos:1980zd, Fei:2016sgs} are other interesting venues for exploration. Finally, gauge symmetries may also be an emergent description in some energy domain of some asymmetric (perhaps entirely unconventional) theory \cite{Iliopoulos:1980zd,Wetterich:2016qee}.

In what follows, we discuss a few examples that illustrate our proposition. While these examples are simplistic and are set for illustration purposes only,  we hope different emergent symmetries discussed below can be incorporated into more elaborate realistic physics models.

 \begin{figure*}[t]
\begin{center}
\includegraphics[scale=0.6]{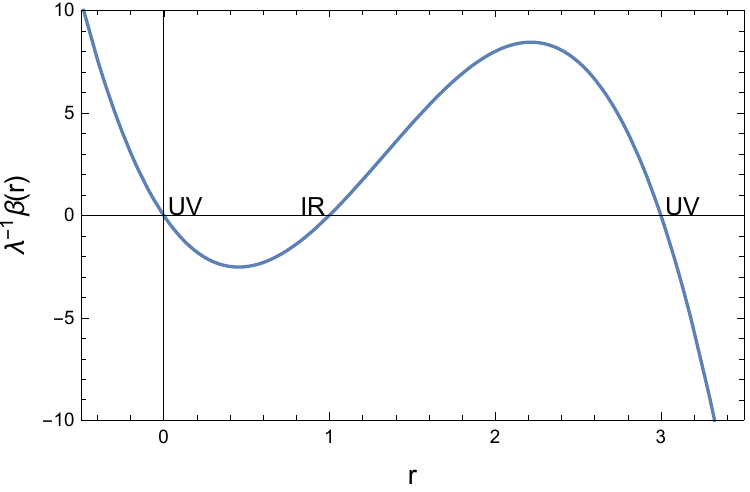}\qquad\qquad
\includegraphics[scale=0.6]{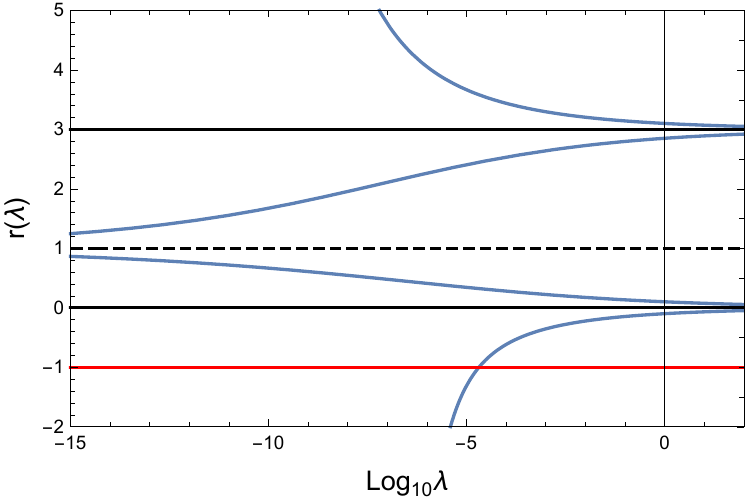}
\end{center}
\caption{Left: Beta function for the ratio of couplings $r=\lambda'/\lambda$. Right: Running for the ratio for different initial conditions. The black solid lines correspond to the decoupling limit, while the dashed one to the enhanced symmetry line. The stability line is also represented in red.}
\label{fglob}
\end{figure*}

\section{Emergent global symmetries}

It is commonly believed that global symmetries are incompatible with a  theory of quantum gravity.
Nevertheless, some global symmetries, such as the chiral symmetry of QCD, are instrumental in understanding low energy physics. It is conceivable to think that such global symmetries are emergent. \\

As a simple example, let us consider first a toy model discussed in \cite{Iliopoulos:1980zd}
which involves  two real scalar fields $\phi_i(x)$  ($i=1,2$).
It exhibits two discrete  symmetries under independent sign reflections
$\phi_1 \to - \phi_1$ and $\phi_2 \to - \phi_2$,
and in addition a discrete exchange symmetry $\phi_1 \leftrightarrow \phi_2$.
The most general Lagrangian including
renormalizable interactions that respect these symmetries reads:\footnote{
We omit the mass terms (relevant operators) which play somewhat passive role in our examples
as well as an ultraviolet cut-off scale $\Lambda$ which may enter the Lagrangian
via non-renormalizable terms (irrelevant operators).
Since we are interested in symmetries emerging in the asymptotics of the RG running
within a particular set of fields, the mass term and cut-off scale serve as infrared (IR) and ultraviolet (UV) asymptotic scales, respectively, i.e. we are dealing with the RG flow for the scales
within the range $m\ll \mu \ll \Lambda$.
 Hence, we shall not consider dimensionful parameters explicitly in what follows.}
\begin{eqnarray}\label{1a}
\mathcal{L}(\phi_i)& =& \frac{1}{2}\left(\partial_{\mu} \phi_i\right)^2  - \mathcal{V}(\phi_i),   \nonumber \\
  \mathcal{V}(\phi_i) &=& \frac{\lambda}{2} \left(\phi_1^4+\phi_2^4\right)
  + \lambda'\,\phi_1^2\,\phi_2^2 \nonumber \\
& =&  \frac{\lambda}{2} \left(\phi_1^2+\phi_2^2\right)^2
 +  (\lambda'-\lambda) \,\phi_1^2\,\phi_2^2
\end{eqnarray}
Stability conditions require $\lambda > 0$ and $\lambda'> -\lambda$.
One can easily see that there are particular adjustments of couplings
which lead to enhanced symmetries in this theory.
Namely, in the limit $\lambda'=\lambda$, the global $O(2)$ symmetry emerges,
whereas $\lambda' =0$ (and $\lambda'=3 \lambda$) corresponds to the decoupling limit
where the original Fock space branches out into two orthogonal Fock spaces
with the associated doubling of symmetries (e.g., Poincare invariance).
For  $\lambda'=3\,\lambda$ the decoupling becomes manifest after changing the field variables,
$\varphi_{1,2} = (\phi_1 \pm \phi_2)/\sqrt2$, so that \eqref{1a} can be rewritten as
\begin{eqnarray}
\mathcal{V}(\phi_i) & \to & \mathcal{V}(\varphi_i) =
 \frac{\lambda+\lambda'}{4} \left(\varphi_1^4+\varphi_2^4\right) +
 \frac{3\lambda-\lambda'}{2} \varphi_1^2\varphi_2^2   \nonumber \\
& = & \frac{\lambda+\lambda'}{4} \left(\varphi_1^2+\varphi_2^2\right)^2   
 + (\lambda \!-\!\lambda') \varphi_1^2\varphi_2^2
\end{eqnarray}
So, Lagrangian $\mathcal{L}(\phi_i)$ can be mapped to the equivalent theory
with $\lambda \to (\lambda+\lambda')/2$ and $\lambda' \to  (3\lambda - \lambda')/2$.
The stability conditions require $\lambda' + \lambda>0$ and $\lambda > 0$.

These limits of the theory are indeed seen in RG flows, by our proposition.
The (one-loop) beta functions read:\footnote{The one-loop factor
$(4\pi)^2$ is absorbed in $\beta_{X}$ so that the corresponding RG
equation reads: $dX/dt=\beta_{X}/(4\pi)^2$, where $t=\ln\mu$.
In some cases we shall need to include two-loop contributions.
Correspondingly, the beta functions including are  presented as
$\beta_X = \beta_X^{(1)} +  \beta_X^{(2)}/(4\pi)^2$.    }
\begin{equation}\label{2a-prime}
\beta_{\lambda}  = 36 \lambda^2+4 \lambda^{\prime2} , \quad\quad
\beta_{\lambda'}  = 24 \lambda \lambda' + 16 \lambda^{\prime2}
\end{equation}
%
%
%
For our purpose, it is sufficient to inspect the RG flow of the ratio of couplings,
$r=\lambda'/\lambda$, which is governed by the following equation:
\begin{eqnarray} \label{3a}
\frac{dr}{dt}&=&\frac{\beta_r^{(1)}}{\left(4\,\pi \right)^2} =
\frac{1}{\left(4\,\pi \right)^2} \frac{\lambda\,\beta^{(1)}_{\lambda'}-\lambda'\,\beta^{(1)}_{\lambda} }{\lambda^2} \nonumber \\
&=& \frac{\lambda}{4\,\pi^2}  \, r \,\left(r-1\right)\,\left(3-r\right) \,.
\end{eqnarray}
It is evident from Eq. (\ref{3a}) that  $r =0,1$ and $3$ are the RG fixed-points.
The nature of these fixed points can be determined by inspecting the derivative of $\beta_{r}$
%
%
near the respective fixed point.
Considering the bounded from below potentials only
 ($\lambda > 0$ and $r> -1$),  we see that $O(2)$ symmetric fixed-line $r=1$ (i.e. $\lambda'=\lambda$)  is an infrared (IR) fixed point; 
  on the contrary, near the fixed points $r=0$ (i.e. $\lambda'=0$)
 and $r=3$ (i.e. $\lambda'= 3\,\lambda$) $d\beta_r/dr$ is negative
 and hence these are UV fixed points, see Fig. \ref{fglob} on the left. Now let us look at Fig. \ref{fglob} on the right: these are the solutions (for different initial conditions) of the equation
 \begin{equation}
     \frac{dr}{dx}=\frac{r\,(r-1)\,(3-r)}{r^2+9}\,,
 \end{equation}
 where $x=\ln\lambda$. This can be obtained by eliminating the $\ln\mu$ dependence from the RG equations.\\
 As we can see, for $1\le r\le 3$ at renormalization scale the theory flows towards the decoupling limit in the UV, while it reassembles itself in a global $O(2)$ symmetric theory in the IR, being $r=1$ the enhanced symmetry line. Moreover, we see that for $r>3$ or $r<0$ at renormalization scale the theory still flows towards the decoupling limit, but the IR fixed lines will be no more an attractor; in the latter case we see also that the theory is no more stable, since the stability condition $r>-1$ will be sooner or later violated.\\

 \begin{figure*}[t]
\begin{center}
\includegraphics[scale=0.6]{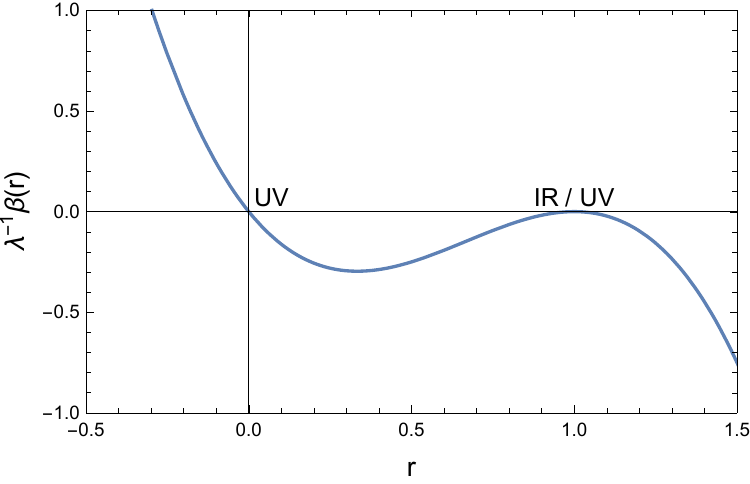}\qquad\qquad
\includegraphics[scale=0.6]{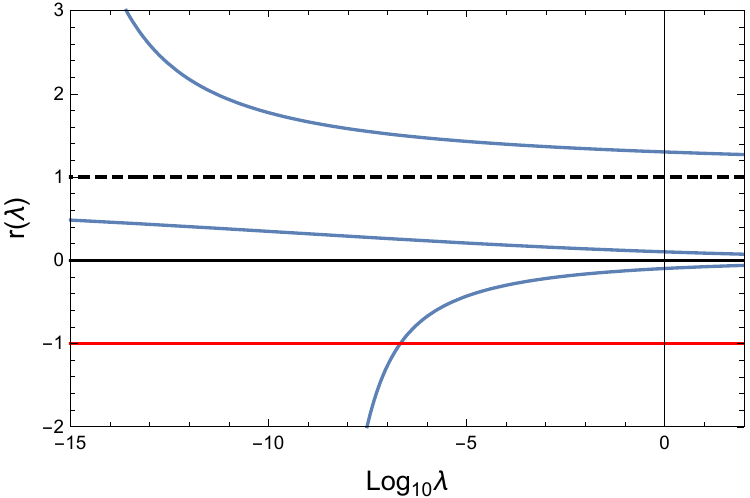}
\end{center}
\caption{Left: Beta function for the ratio of couplings $r=\lambda'/\lambda$. Right: Running for the ratio for different initial conditions. The black solid line corresponds to the decoupling limit, while the dashed one to the enhanced symmetry line. The stability line is also represented in red.}
\label{2complex}
\end{figure*}

As a straightforward generalization, let us consider
a model with 2 complex scalars $\phi_1$ and $\phi_2$.
The most general potential consistent with a $U(1)_1\times U(1)_2$ symmetry
together with
discrete exchange symmetry $\phi_{1} \leftrightarrow \phi_{2}$ is
\begin{eqnarray}\label{1a}
\mathcal{V}&=&  
  \frac{\lambda}{2} \,(\phi_1^\dagger \phi_1)^2 +
\frac{\lambda}{2} \,(\phi_2^\dagger \phi_2)^2
+ \lambda' \,(\phi_1^\dagger \phi_1)  (\phi_2^\dagger\phi_2) \nonumber \\
& =&  \frac{\lambda}{2}\, \left(\vert\phi_1\vert^2 + \vert\phi_2\vert^2 \right)^2
 +  (\lambda'-\lambda)\, \vert \phi_1\vert^2 \vert\phi_2\vert^2
\end{eqnarray}
The one and two loop beta functions for these couplings read
\begin{align}
& \beta^{(1)}_{\lambda} = 10\,\lambda^2+2\,\lambda^{\prime2} , \quad
\beta^{(2)}_{\lambda} = -60\,\lambda^3 -10 \,\lambda \,\lambda^{\prime2}
-8\, \lambda^{\prime3} \nonumber \\
& \beta^{(1)}_{\lambda'} = 8\, \lambda\lambda' + 4\,\lambda^{\prime2} , \quad
\beta^{(2)}_{\lambda'} = -20\,\lambda^2\,\lambda' -48 \,\lambda \,\lambda^{\prime2}
-10 \,\lambda^{\prime3} \nonumber
\end{align}
The RG flow for the ratio $r=\lambda'/\lambda$ is governed by the one loop equation
\begin{equation}
    \frac{dr}{dt}=-\frac{\lambda}{8\,\pi^2}\,r\,(r-1)^2\,.
\end{equation}
It is clear that the fixed points are $r=0$ (corresponding to the decoupling limit) and $r=1$ (corresponding to an enhanced global $U(2)$ symmetry). Although $r=0$ results to be an UV fixed point (since the derivative of $\beta^{(1)}_r$ is negative) we need for a two loop contribution in order to determine the nature of the other fixed point, since it gives a null derivative at one loop\footnote{Formally one can see that, for $0<r<1$ at renormalization scale, the theory reassembles itself into an $U(2)$ symmetric theory in the IR and flows towards the decoupling limit in the UV, while for $r>1$ the theory is decoupled in the UV, but there is no IR fixed line.}; in Fig. \ref{2complex} we show the analogous results of the previous case. A simple calculation provides
\begin{equation}
    \beta^{(2)}_r=8\,\lambda^2\,r\,(r-1)\,(r^2+r-5)\,,
\end{equation}
whose derivative is negative for $r=1$, thus being an UV fixed point. Moreover, $d\beta^{(2)}_r/dr|_{r=0}>0$: one can notice that the two-loop contribution has the opposite behaviour with respect to the one loop case. At the perturbative level, this will not change the nature of the fixed point since this contribution will be very suppressed by the two loop phase space extra factor $16\,\pi^2$ with respect to the previous one\footnote{Indeed, the derivative of the full beta $\beta_r^{(1)}+\frac{1}{16\,\pi^2}\,\beta_r^{(2)}$ is still negative for $r=0$.}.\\

\begin{figure}[t]
\includegraphics[scale=0.6]{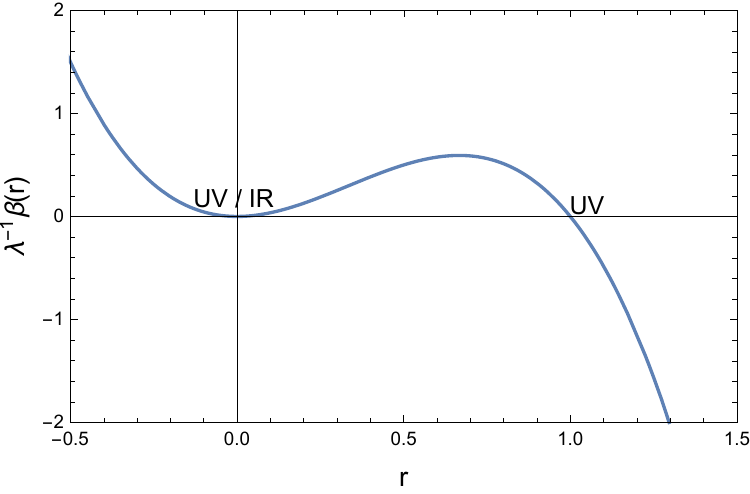}
\caption{One loop beta function for the ratio of couplings for the 2 Higgs model. On the local minimum it is not possible to determine the nature of the fixed point; a higher loop level is needed.}
\label{2hbr}
\end{figure}

As a next example,
let us consider two complex doublets $\phi_1$ and $\phi_2$
of two different global symmetries  $U(2)_1$ and $U(2)_2$.
The general Lagrangian includes the terms
\begin{eqnarray}\label{2-higgs}
\mathcal{V} &=&
-  \frac{\lambda}{2} \,(\phi_1^\dagger \phi_1)^2 -
\frac{\lambda}{2} \,(\phi_2^\dagger \phi_2)^2
- \lambda' \,(\phi_1^\dagger \phi_1)  (\phi_2^\dagger\phi_2) \nonumber \\
& =& - \frac{\lambda}{2}\, \left(\vert\phi_1\vert^2 + \vert\phi_2\vert^2 \right)^2
 -  (\lambda'-\lambda)\, \vert \phi_1\vert^2 \vert\phi_2\vert^2
\end{eqnarray}
where we again impose a discrete symmetry $\phi_1\leftrightarrow \phi_2$
under the exchange of two $U(2)$ factors.  Obviously, in the limit
$\lambda'-\lambda \to 0$ this model acquires a larger symmetry $U(4)$.
The one and two loop beta functions are
\begin{align}\label{beta-2higgs-lambdapr}
& \beta^{(1)}_{\lambda} = 12\,\lambda^2+4\,\lambda^{\prime2} , \quad
\beta^{(2)}_{\lambda} = -78\,\lambda^3 -20 \,\lambda \,\lambda^{\prime2}
-16\, \lambda^{\prime3} \nonumber \\
& \beta^{(1)}_{\lambda'} = 12\, \lambda\lambda' + 4\,\lambda^{\prime2} , \quad
\beta^{(2)}_{\lambda'} = -72\,\lambda^2\,\lambda' -30 \,\lambda \,\lambda^{\prime2}
-12 \,\lambda^{\prime3} \nonumber
\end{align}
%
%
%
We can determine the nature of the fixed points by inspecting the RG flow of the ratio $r=\lambda'/\lambda$; at one loop level we get
\begin{equation}
    \beta^{(1)}_r=-4\,\lambda\,r^2\,(r-1)
\end{equation}
which has $r=0$ and $r=1$ as fixed points; the former corresponds to the decoupling limit.\\
By evaluating the derivative of this beta
\begin{equation}
    \frac{d\beta_r^{(1)}}{dr}=4\,\lambda\,r\,(2-3\,r)
\end{equation}
we see that it is negative for $r=1$, being it an UV fixed point, while it is zero for $r=0$; this means that we need to inspect the two loop corrections in order to determine the nature of the fixed point relative to the decoupling limit, see Fig. \ref{2hbr}. Indeed
\begin{equation}
    \beta^{(2)}_r=2\,\lambda^2\,r\,(r-1)\,(8\,r^2+12\,r-3)\,.
\end{equation}
Evaluating the derivative of this beta with respect to $r$ for $r=0$ we get $d \beta^{(2)}_r/dr|_{r=0}>0$; it is an IR fixed point. In summary, there is an UV attractive fixed point corresponding to an emergent global $U(4)$ symmetry.

\section{Emergent supersymmetry}
Supersymmetry is the unique non-trivial extension of Poincare invariance, which provides important insights into several theoretical and phenomenological aspects
of high energy physics, such as strongly coupled theories or stability of hierarchies of scales.
Therefore, it is interesting to study systems that exhibit emergent supersymmetry.

\subsection{Emergent Wess-Zumino model}

A theory, for having a possible supersymmetric realization, must contain an
equal amount of fermionic and bosonic degrees of freedom.
A simple system that exhibits  this property 
is a Higgs-Yukawa model with a complex scalar field $\phi$ and a 2-component Weyl spinor $\psi$. The most generic Lagrangian containing only renormalizable interactions reads:\footnote{For the sake
of brevity, we choose to enforce an unbroken global $U(1)$ symmetry:
$\phi\to e^{2\,i\,\alpha}\,\phi,~\psi\to e^{-i\,\alpha}\,\psi$.}
\begin{align}\label{WZ}
\mathcal{L}=|\partial_{\mu}\phi|^2  +i\,\psi^{\dagger}\,\bar \sigma^{\mu}\,\partial_{\mu}\psi
-\!\lambda\,(\phi^\ast \phi)^2  -y (\phi\,\psi^2 + \mathrm{h.c.})
\end{align}
where the Yukawa coupling constant $y$ can be made real by a phase transformation of the fields.
For $\lambda =y^2$ this theory exhibits an enhanced spacetime symmetry,
the (on-shell) ${\cal N}=1$ supersymmetry, when the Lagrangian can be obtained from the superpotential $W=(y/3)\,\Phi^3$ with $\Phi$ being a chiral superfield containing $\phi$ and $\psi$.
We did not include in the Lagrangian (\ref{WZ}) the mass term.
The fermion mass term  $m\,\psi^2$ is forbidden by $U(1)$ symmetry which in supersymmetric
context corresponds to $U(1)_R$ symmetry.  The scalar mass term
$\mu^2 \phi^\dagger\phi$ can be interpreted as a soft supersymmetry breaking term.

\begin{figure}[t]
\includegraphics[scale=0.6]{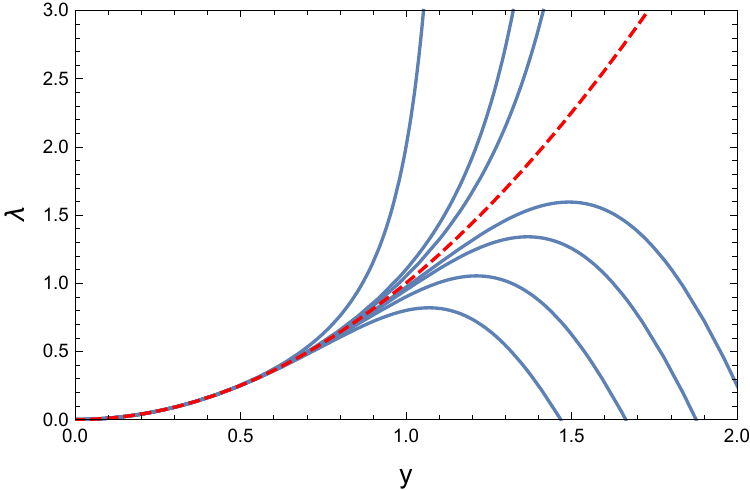}
\caption{RG flows towards $\mathcal{N}=1$ supersymmetric IR fixed-line $\lambda=y^2$ (red dashed) in the Wess-Zumino model.}
\label{f2}
\end{figure}

Let us examine RG flows of couplings, which are governed (at one loop) by the following
$\beta$-functions \cite{Machacek:1983tz}:
\begin{align}\label{WZ-1loop}
\beta_y^{(1)}= 6\,y^3 , \quad \quad
\beta_{\lambda}^{(1)}= 20\,\lambda^2+8\,\lambda\,y^2-16\,y^4 .
\end{align}
As in the previous section, it is sufficient to inspect the evolution of the ratio, $r=\lambda/y^2$.
At  one loop, we get
\begin{equation}\label{r}
\frac{\beta^{(1)}_r}{16\,\pi^2}=\frac{dr}{dt}=\frac{y^2}{4\pi^2} \,  (r-1) (4+5r)\,.
\end{equation}
In the stable domain ($\lambda >0$), the model indeed exhibits ${\cal N}=1$ supersymmetric  fixed-line $\lambda=y^2$ (i.e. $r=1$), together with the decoupling fixed-point $y=0$. By taking the derivative of $\beta_r^{(1)}$ in \eqref{r} we get
\begin{equation}\label{der}
\frac{d\beta_r^{(1)}}{dr}=4\,y^2\,(10\,r-1)\,.
\end{equation}
Evaluating this expression in $r=1$, we get a positive value. Therefore, the supersymmetric fixed-line is an IR attractor (see also Fig. \ref{f2}). \\
Now we calculate also two-loop contributions:
\begin{align}\label{WZ-2loop}
& \beta_y^{(2)}= 4\,y^5 -32\,y^3\,\lambda+4\,y\,\lambda^2  , \nonumber \\
& \beta_{\lambda}^{(2)}= -240\,\lambda^3 - 80 \,\lambda^2 \,y^2 + 16 \,\lambda \,y^4 + 256 \,y^6 .
\end{align}
In the supersymmetric limit $\lambda=y^2 \equiv a$, one has
\begin{equation}\label{susylim}
\beta_\lambda=2 \,y \,\beta_y\,.
\end{equation}
This remains satisfied even at two-loop level: indeed, one obtains a unique beta function
\begin{equation}
\beta^{(2)}_\lambda=2\,y\,\beta^{(2)}_{y}= -48\,a^6 \,.
\end{equation}

The Lagrangian \eqref{WZ} can be presented as
\begin{align}\label{WZ-D}
{\cal L} & = & \int d\theta d\bar\theta \Phi^\dagger \Phi + \left[\int d \theta W(\Phi) + \text{h.c}\right]
\nonumber \\
&& + \frac{1}{M} \int d\theta d\bar\theta \eta \bar\eta (\Phi^\dagger \Phi)^2
\end{align}
where the first two terms
correspond to the supersymmetric Lagrangian of the Wess-Zumino
model with $\Phi(\phi,\psi)$ being a chiral superfield and superpotential $W(\Phi) = y \Phi^3$,
i.e.  with the scalar quartic coupling in Lagrangian  \eqref{WZ}  taken as $\lambda = y^2$.
The second term can be considered as a (hard) supersymmetry breaking $D$-term,
with $\eta = \mu \theta^2$ being a supersymmetry breaking spurion with non zero $F$-term.
Hence, for non-supersymmetric coupling constant
we have $\bar{\lambda} = \lambda - y^2 = \mu^2/M$.

For $\beta$-function of   $\bar{\lambda} = \lambda - y^2$ we get
\begin{align}\label{tilde-lambda}
& \beta^{(1)}_{\bar{\lambda}} = \bar{\lambda}\, (48\, y^2 + 20\,\bar{\lambda}) \\
& \beta^{(2)}_{\bar{\lambda}} = - \bar{\lambda} \,(816 \,y^4 + 808\, \bar{\lambda} \,y^2 + 240 \,\bar{\lambda}^2)
\nonumber
\end{align}
We see that supersymmetry breaking $D$-term in \eqref{WZ-D} disappears at low energies: $\bar{\lambda} =0$ is an IR fixed point 
at two-loop level.

\subsection{Emergent ${\cal N}=1$ supersymmetric gauge theory}

Here we consider a few models with gauge symmetries: first, we focus on a Yang-Mills theory, showing there the possibility of having an IR emergent supersymmetry; second, we emphasize that one cannot achieve the same situation if the starting point is a model with a $U(1)$ gauge symmetry only.\\

\paragraph{$SU(2)$ gauge model.}
Let us take  a pure Yang-Mills model
(with a gauge symmetry e.g. $SU(2)$ for simplicity)
complemented by a two-component fermion  $\chi$ in adjoint (triplet) representation, so that
the number of bosonic and fermionic degrees of freedom are equal.
The Lagrangian of this theory reads:
\begin{equation}\label{pure}
\mathcal{L}_0 = -\frac14 \,G_{a\,\mu\nu} \,G_a^{\mu\nu}
+ i\,\chi^{\dagger}\,\bar \sigma^{\mu} \,D_{\mu}\chi  
\end{equation}
where $D_\mu = \partial_\mu - i\,g \,A^a_\mu\, T_a$ is a covariant derivative,
with $T_a$ ($a=1,2,3$) being generators of $SU(2)$ in respective representation.
Hence, this model contains only a gauge coupling constant $g$, with $\beta_g^{(1)} = -6\,g^3$.

This model automatically exhibits exact ${\cal N}=1$ supersymmetry,
as a consequence of the gauge symmetry:
Lagrangian (\ref{pure}) describes ${\cal N}=1$ super Yang-Mills theory.
The mass term $\frac{m}{2}\big(\chi^2 + {\rm h.c.} \big)$, if any, can be interpreted as a soft supersymmetry breaking term.

Next, we consider a model that includes gauge interactions and multiple couplings. We add matter species as a Weyl fermion $\psi$ and a scalar $\phi$ both in a doublet representation of $SU(2)$,
so that the numbers of fermion and boson degrees of freedom are again equal.
(In fact, such a toy theory is ill-defined since it has global $SU(2)$  anomaly~\cite{Witten:1982fp}
but we consider it first for the sake of simplicity).
The most general interaction Lagrangian, besides the gauge interactions \eqref{pure},
contains the following terms with dimensionless coupling constants:
\begin{eqnarray}\label{one-fl}
\mathcal{L} & =&  -\frac{\lambda}{8} \left(\phi^{\dagger} \phi\right)^2
-\frac{1}{\sqrt{2}}\left( y \, \phi^\dagger \,\tau^a \,\psi\, \chi^a + \mathrm{h.c.}\right)
\end{eqnarray}
where $\tau^a$ ($a=1,2,3$) are the Pauli matrices which act, in the second term, between the doublets $\phi$ and $\psi$.
The Yukawa constant $y$ can be rendered real and positive by the phase transformations.
The vacuum stability condition implies $\lambda > 0$ for the quartic scalar coupling.

 \begin{figure*}[t]
\begin{center}
\includegraphics[scale=0.8]{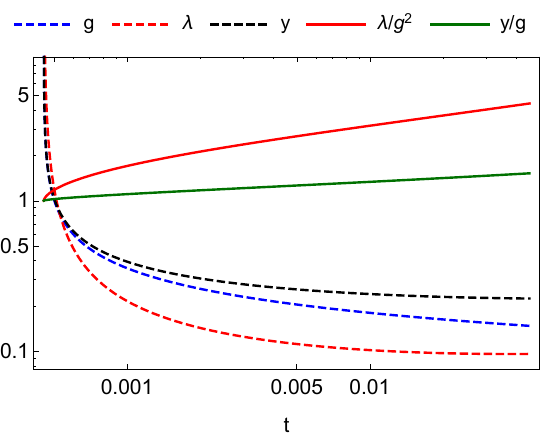}\qquad\qquad
\includegraphics[scale=0.65]{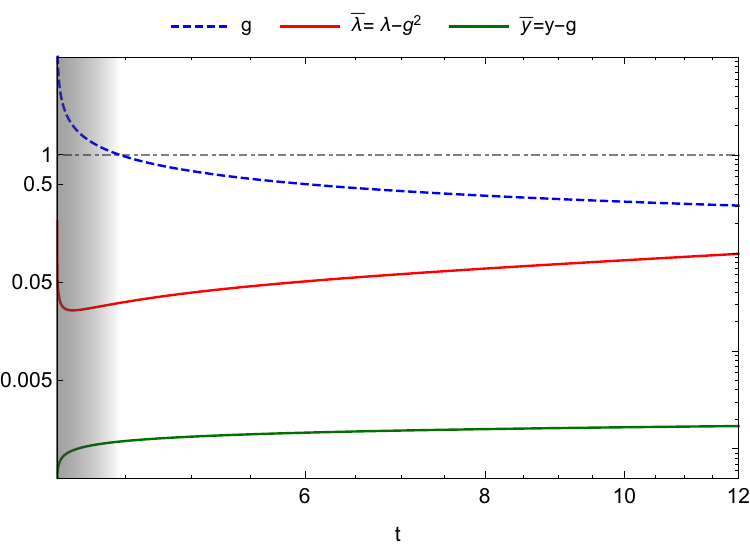}
\end{center}
\caption{Left: IR convergence of the ratios of coupling constants of Lagrangian \eqref{one-fl}. Right: Running for both $\bar{\lambda}=\lambda-g^2$ and $\bar{y}=y-g$ along the perturbative range. The coupling $\bar{\lambda}$ starts to grow only in the perturbatively unreliable zone, denoted by the shaded area.}
\label{f3}
\end{figure*}

The (one-loop) beta functions of this model read:
\begin{eqnarray}
&& \beta_g  = -\frac{11}{2}\,g^3\, ,  \quad \quad \beta_y  = \frac{11}{4}\,y^3-\frac{33}{4}\,g^2\, y
\nonumber \\
&& \beta_\lambda = 9\,g^4 - 9\,g^2\, \lambda  -20\, y^4  + 6\, y^2  \,\lambda + 3 \,\lambda^2
\end{eqnarray}
So beta functions of ratios  $\bar y \equiv y/g$ and $\bar \lambda \equiv \lambda/g^2$ are
\begin{eqnarray}
\beta_{\bar y} &=&\frac{11 }{4}\, g^2\,  \bar y\left(\bar y^2-1\right) \nonumber \\
\beta_{\bar \lambda}  &=& g^2 \left(3\,\bar\lambda^2
+2\,\bar\lambda + 6\,\bar\lambda \,\bar{y}^2- 20 \,\bar y^4+ 9 \right)
\end{eqnarray}
which exhibit the following fixed-line:
\begin{eqnarray}\label{one-fl-fix}
\vert \bar y \vert = \bar \lambda = 1
\end{eqnarray}
It can be readily checked that this fixed line is an IR stable.
The RG evolution is shown on the left of Fig.~\ref{f3}.   \\

 \begin{figure*}
\begin{center}
\includegraphics[scale=0.6]{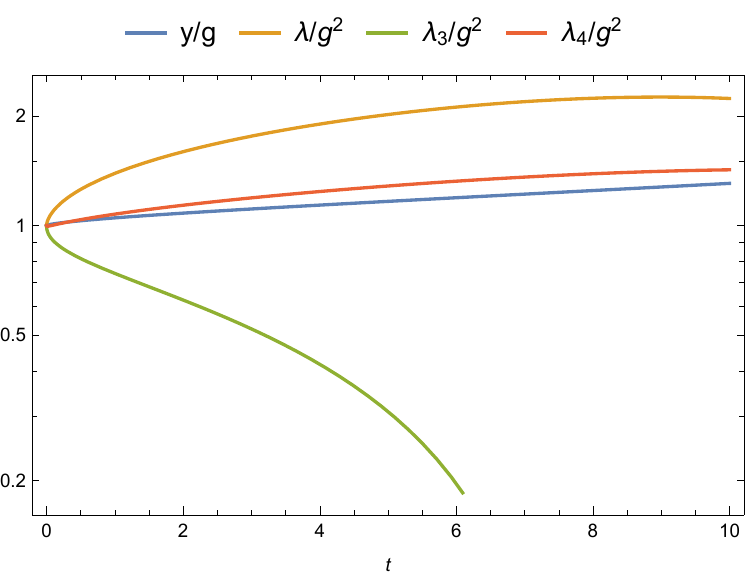}\qquad
\includegraphics[scale=0.6]{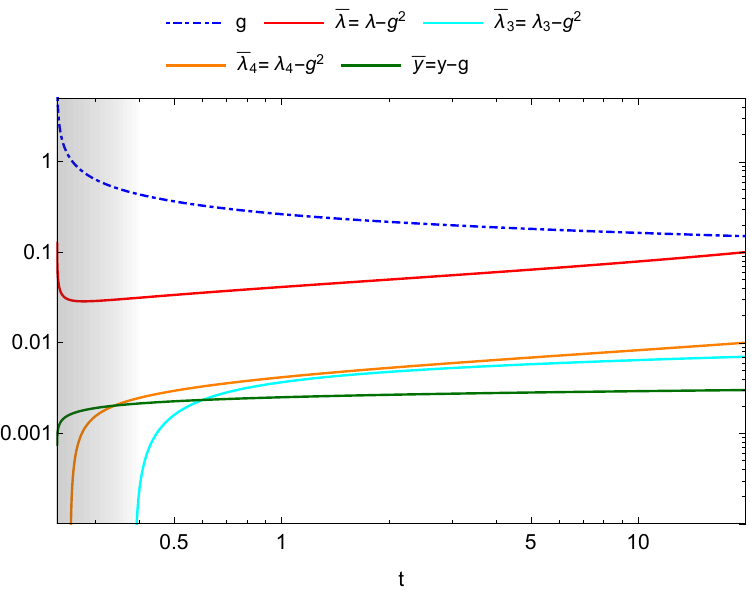}
\end{center}
\caption{Left: Running for the ratios of coupling constants of Lagrangian (\ref{2HDM}). Right: Running for the differences. The shaded area denotes the perturbatively unreliable zone.}
\label{f4}
\end{figure*}

Equivalently, beta functions of the differences
$\tilde{y} \equiv y-g$ and $\tilde{\lambda} \equiv \lambda-g^2$ are
\begin{align}
 \beta_{\tilde{y}} = & \frac{11 }{4}\, \tilde{y}^2 (3\,g + \tilde{y}) \nonumber \\
 \beta_{\tilde \lambda} = &\, 3\,\tilde{\lambda}\,(g^2 + 4\,g\,\tilde{y} + 2\,\tilde{y}^2 + \tilde{\lambda}^2 )
  \nonumber \\
&  - 2\,\tilde{y}\, (34\,g^3 + 57\, g^2\, \tilde{y} +40 \,g\,\tilde{y}^2 +10 \,\tilde{y}^3)  \,.
\end{align}
It follows that the values $\tilde{y}=\tilde{\lambda}=0$ are zeros of the betas.
Notice that the determinant of the matrix
\begin{equation}
    M_{i j}:= \frac{d \beta^{(1)}_{g_i}}{dg_j} \,\,\,\,\,\,\, \text{with} \,\,\,\,\,\,\,\,\,  g_i=(\bar{y},\bar{\lambda})
\end{equation}
is zero on that fixed point. Therefore, at this level this point is not attractive, nor repulsive. The behavior of these couplings is on the right of Fig. \ref{f3}.
By the way, this is strongly dependent on the initial conditions for the renormalization group equations, in contrast with what happens for the ratios discussed above.\\

Using the Fierz identities for the Pauli matrices,
Lagrangian (\ref{one-fl}) can be rewritten as
\begin{eqnarray}\label{one-fl-D}
\mathcal{L} =- \frac{\bar\lambda\,g^2}{8} (\phi^{\dagger} \tau^a\phi)^2
- \frac{\bar y \,g}{\sqrt{2}}\left(\phi^\dagger\,\tau^a\, \psi\,  \chi^a  + \mathrm{h.c.}\right) \,,
\end{eqnarray}
which shows that in the limit \eqref{one-fl-fix}, i.e.  $y=g,~\lambda =g^2$, the theory tends to (on-shell) $\mathcal{N}=1$ supersymmetric $SU(2)$ gauge theory of a vector supermultiplet $(A^a_{\mu}, \chi^a)$ and  chiral supermultiplet $(\phi, \psi)$
interacting via supergauge interactions with a coupling constant $g$.

\medskip

Let now us generalize what discussed above by considering a model
with a pair of doublets $(\phi_i, \psi_i)$, $i=1,2$.
The Yukawa couplings, without losing generality, can be taken diagonal
\begin{align}\label{2HDM-Yuk}
-\frac{y_1}{\sqrt2} \,\phi_1^\dagger\, \tau^a \,\psi_1 \,\chi^a
- \frac{y_2}{\sqrt2} \, \phi_2^\dagger \,\tau^a \,\psi_2 \,\chi^a   + \mathrm{h.c.}
\end{align}
with constants $y_{1,2}$ being real and positive.

The most general scalar potential reads
\begin{align}\label{2HDM}
{\cal V}(\phi_1,\phi_2) & = \frac{\lambda_1}{8} \,(\phi_1^{\dagger} \phi_1)^2 +
\frac{\lambda_2}{8} \,(\phi_2^{\dagger} \phi_2)^2 - \frac{\lambda_3}{4}\, (\phi_1^{\dagger} \phi_1)(\phi_2^{\dagger} \phi_2)
\nonumber \\
&
+ \frac{\lambda_4}{2} \,(\phi_1^{\dagger} \phi_2)(\phi_2^{\dagger} \phi_1)   + \frac{1}{4} \left[\lambda_5\, (\phi_1^\dagger \phi_2)^2 \right.
 \nonumber  \\
&  \left. +\lambda_6 \,(\phi_1^\dagger \phi_1)(\phi_1^\dagger \phi_2) +\lambda_7\, (\phi_2^\dagger \phi_2)(\phi_2^\dagger \phi_1)+ \text{h.c.} \right]
 \end{align}
 The constant $\lambda_5$ can be taken real; moreover, for the sake of simplicity, we can impose a discrete sign change symmetry $\phi_{1,2}\leftrightarrow-\phi_{1,2}$ together with the exchange symmetry $\phi_1\leftrightarrow\phi_2$, to set $\lambda_1=\lambda_2=\lambda$, $\lambda_6=\lambda_7=0$ and $y_1=y_2=y$ in the previous Lagrangians. These further assumptions will not affect our considerations.

Using Fierz identities for Pauli matrices, we obtain
\begin{align}\label{Fierz}
& (\phi_1^\dagger \tau^a\phi_1)(\phi_2^\dagger \tau^a\phi_2) =
\nonumber \\
& = \frac32  (\phi_1^\dagger \phi_2)(\phi_2^\dagger \phi_1) -
\frac12  (\phi_1^\dagger \tau^a\phi_2)(\phi_2^\dagger \tau^a\phi_1)
\nonumber \\
& = 2(\phi_1^\dagger \phi_2)(\phi_2^\dagger \phi_1) - (\phi_1^\dagger \phi_1)(\phi_2^\dagger \phi_2)
\end{align}
Then the scalar potential can be rewritten as
\begin{align}\label{2HDM-D}
 & {\cal V}(\phi_1,\phi_2) =  \frac{g^2}{2} D_a^2 + \frac{\tilde\lambda}{8} (\phi_1^{\dagger} \phi_1)^2 +
\frac{\tilde\lambda}{8} (\phi_2^{\dagger} \phi_2)^2 +
\nonumber \\
&  - \frac{\tilde\lambda_3}{4} (\phi_1^{\dagger} \phi_1)(\phi_2^{\dagger} \phi_2)
+ \frac{\tilde\lambda_4}{2} (\phi_1^{\dagger} \phi_2)(\phi_2^{\dagger} \phi_1)
 \nonumber  \\
&  +  \frac{\lambda_5}{4} \left[ (\phi_1^{\dagger} \phi_2)^2   + \mathrm{h.c.} \right]
  \end{align}
where $\tilde\lambda_i = \lambda_i-g^2$ are in fact constants of effective supersymmetry
breaking terms.

 One-loop beta functions of this theory are
\begin{align}\label{RG-gy12}
& \beta_g = -5\,g^3  \\
& \beta_y = -\frac{33}{4}\,g^2\, y + \frac{13}{4}\,y^3
\end{align}
and
\begin{align}\label{RG-lambda}
 \beta_\lambda  =&\, 9\,g^4 - 9 \,g^2\, \lambda -20 \,y^4  + 6 \,y^2 \, \lambda     \nonumber \\
  & + 3\, \lambda^2 + \lambda_3^2 - 2\, \lambda_3\,\lambda_4 + 2 \,\lambda_4^2  + 2\, \lambda_5^2  \\
 \beta_{\lambda_3}  = &   -9\,g^4  - 9 \,g^2\, \lambda_3 + 4 \,y^4 + 6\,y^2\, \lambda_3
 \nonumber \\
&   - \lambda_3^2 +  \lambda \,(3\, \lambda_3 -2\,\lambda_4 )
   - 2 \,\lambda_4^2 -  2 \,\lambda_5^2   \\
\beta_{\lambda_4}  =& - 9\,g^2\, \lambda_4 - 8\,y^4 + 6 \,y^2\, \lambda_4
\nonumber \\
& + 2\,\lambda_4^2 +  \lambda\, \lambda_4 - 2\,\lambda_3 \,\lambda_4 + 4\,\lambda_5^2  \\
\beta_{\lambda_5}  =& \lambda_5\,\left(6\,y^2 - 9\,g^2  + \lambda - 2\,\lambda_3  + 6\,\lambda_4 \right) \,.
\end{align}
Defining the ratios $\bar{y}=y/g$, $\bar{\lambda}_i=\lambda_i/ g^2$, the relative beta functions are
\begin{align}
\beta_{\bar{y}} = &\frac{13}{4}\,g^2 \,\bar{y}\,\left(\bar{y}^2-1\right) \\
\beta_{\bar{\lambda}} =& g^2 \,\left(\bar{\lambda}_3^2-2 \,\bar{\lambda}_3\, \bar{\lambda}_4+2\,\bar{\lambda}_4^2+3\, \bar{\lambda}^2+\bar{\lambda} \right.  \nonumber \\
& \left. -20\, \bar{y}^4+6 \,\bar{\lambda}\, \bar{y}^2+9\right)  \\
\beta_{\bar{\lambda}_3} =& g^2\, \left(-\bar{\lambda}_3^2+3\, \bar{\lambda}_3 \,\bar{\lambda}+\bar{\lambda}_3-2\, \bar{\lambda}_4 \,(\bar{\lambda}_4+\bar{\lambda})  \right. \nonumber \\
&\left. +4 \,\bar{y}^4+6\,\bar{\lambda}_3\, \bar{y}^2-9\right)  \\
\beta_{\bar{\lambda}_4} =& g^2 \,\left(\bar{\lambda}_4 \,(-2\, \bar{\lambda}_3+2\, \bar{\lambda}_4+\bar{\lambda}+1)\right.\\
&\left.-8\, \bar{y}^4+6 \,\bar{\lambda}_4 \,\bar{y}^2\right)\,,
\end{align}
which are null for $\bar{y} = \bar{\lambda}_i=1$. The derivatives
of $(\beta_{\bar{y}}, \beta_{\bar{\lambda}}, \beta_{\bar{\lambda}_3}, \beta_{\bar{\lambda}_4})$ with respect of $(\bar{y}, \bar{\lambda}, \bar{\lambda}_3, \bar{\lambda}_4)$ at the fixed point reads
\begin{equation}
\left(
\begin{array}{cccc}
 13\,g^2/2 & 0 & 0 & 0 \\
 -68 \,g^2 & 13 \,g^2 & 0 & 2 \,g^2 \\
 28\, g^2 & g^2 & 8 \,g^2 & -6 \,g^2 \\
 -20\, g^2 & g^2 & -2 \,g^2 & 10\, g^2 \\
\end{array}
\right)\,,
\end{equation}
which has positive determinant, meaning that the fixed point is IR attractive.

We show in Fig. \ref{f4} (on the left) the running for the ratios. Moreover, in complete analogy with the discussion on the model with less coupling in \eqref{one-fl}, we also plot the differences $y-g$, $\lambda_i - g^2$ ($i=1,2,3,4$) in Fig. \ref{f4} (on the right).   \\


Let us consider now the case of abelian $U(1)$ gauge symmetry.
As anticipated in the beginning of this subsection, the case of an abelian
gauge theory does not share with the Yang-Mills case the feature of having an attractive fixed point, corresponding to an enhanced supersymmetry. In what follows, we explicitly show this point.

Consider a $U(1)$ gauge-invariant theory with a pair of two-component fermions $\psi_1$ and $\psi_2$, a pair of scalar fields $\phi_1$ and $\phi_2$ which carry $U(1)$ charges $q_{\psi_1}=q_{\phi_1}=-q_{\psi_2}=-q_{\phi_2}=1$ and a zero-charge two-component fermion $\xi$.
The interaction Lagrangian of the model reads:
\begin{align}
\mathcal{L} =& - \left( \frac{\lambda_1}{2} |\phi_1|^4 + \frac{\lambda_2}{2} |\phi_2|^4 + \lambda_3 |\phi_1|^2 |\phi_2|^2 \right) \nonumber \\
& - \sqrt{2}  \left( y_1 \,\xi\, \psi_1\, \phi_1^* +  y_2\, \xi \,\psi_2 \,\phi_2^* \right) \nonumber \\
& + \text{gauge interaction terms}
\end{align}
The one-loop beta functions for this model are:
\begin{align}
 \beta_g^{(1)} =  \,&        2 \,g^3                                               \\
 \beta_{y_{1}}^{(1)} =  &\,  y_1 \left(-3 \,g^2+4 \,y_1^2+y_2^2\right)                                               \\
 \beta_{y_{2}}^{(1)} = &\,   y_2 \left(-3 \,g^2+y_1^2+4\, y_2^2\right)                                               \\
 \beta_{\lambda_{1}}^{(1)} =&\,   2 \left(6\, g^4-6\, g^2\, \lambda_1+5\, \lambda_1^2+\lambda_3^2\right.\\
 &\left.-8 \,y_1^4+4 \,\lambda_1 \,y_1^2\right)\\
 \beta_{\lambda_{2}}^{(1)} =&\,  2 \left(6 \,g^4-6 \,g^2\, \lambda_2+5 \,\lambda_2^2+\lambda_3^2\right.\\
 &\left.-8 \,y_2^4+4 \,\lambda_2 \,y_2^2\right)\\
 \beta_{\lambda_{3}}^{(1)} = &\, 4 \left(3 \,g^4-4\, y_1^2 \,y_2^2\right)+4 \,\lambda_3 \left(-3 \,g^2+\lambda_1\right.  \nonumber \\
&\left.+\lambda_2+y_1^2+y_2^2\right)+4 \,\lambda_3^2
\end{align}
We focus on the running of ratios of couplings:  $\bar{y}_1= y_1/g$, $\bar{y}_2= y_2/g$, $\bar{\lambda}_1= \lambda_1/g^2$, $\bar{\lambda}_2= \lambda_2/g^2$, $\bar{\lambda}_3= \lambda_3/g^2$, whose beta functions read
\begin{align}
 \beta^{(1)}_{\bar{y}_{1}} = &\,  g^2\, \bar{y}_1 \left(4 \,\bar{y}_1^2+\bar{y}_2^2-5\right)                                       \\
 \beta^{(1)}_{\bar{y}_{2}} = &\, g^2\, \bar{y}_2 \left(\bar{y}_1^2+4 \,\bar{y}_2^2-5\right)                                        \\
 \beta^{(1)}_{\bar{\lambda}_{1}} = &\, 2 \,g^2 \left(5\, \bar{\lambda}_1^2-8 \,\bar{\lambda}_1+\bar{\lambda}_3^2-8 \,\bar{y}_1^4+4\, \bar{\lambda}_1\, \bar{y}_1^2+6\right)                            \\
 \beta^{(1)}_{\bar{\lambda}_{2}} =&\, 2 \,g^2 \left(5 \,\bar{\lambda}_2^2-8 \,\bar{\lambda}_2+\bar{\lambda}_3^2-8 \,\bar{y}_2^4+4\,\bar{\lambda}_2 \,\bar{y}_2^2+6\right)                            \\
 \beta^{(1)}_{\bar{\lambda}_{3}} =&\, 4 \,g^2 \left(\bar{\lambda}_3^2+\bar{y}_1^2 \left(\bar{\lambda}_3-4\, \bar{y}_2^2\right)  \right.   \nonumber \\
&\left. +\bar{\lambda}_3 \left(\bar{\lambda}_1+\bar{\lambda}_2+\bar{y}_2^2-4\right)+3\right)
\end{align}
The possible zeros are
\begin{align}
& |\bar{y}_1 | = |\bar{y}_2 | = 1\,, \,\,\,\,\,\,   \bar{\lambda}_1 = \bar{\lambda}_2= |\bar{\lambda}_3|= 1               \label{case1}                   \\
& |\bar{y}_1 | = |\bar{y}_2 | = 1\,, \,\,\,\,\,\,   \bar{\lambda}_1 = \bar{\lambda}_2= \frac{7}{15}\,, \,\,\,\, \bar{\lambda}_3 =  \frac{5}{3}   \label{case2}
\end{align}
Consider now the matrix of the derivatives
$$
M_{i j}:= \frac{d \beta^{(1)}_{g_i}}{dg_j} \,\,\,\,\,\,\, \text{with} \,\,\,\,\,\,\,\,\,  g_i=(\bar{y}_1, \bar{y}_2, \bar{\lambda}_1, \bar{\lambda}_2, \bar{\lambda}_3)   \,.
$$
Eq. \eqref{case2} gives negative eigenvalues for $M_{i j}$ (so IR unstable).

Inside Eq. \eqref{case1}, the only subcase giving positive eigenvalues for $M_{i j}$ (
thus being a IR stable fixed point) is
\begin{equation}
|\bar{y}_1 | = |\bar{y}_2 | = 1\,, \,\,\,\,\,\,   \bar{\lambda}_1 = \bar{\lambda}_2= \bar{\lambda}_3= 1 \,,
\end{equation}
while the eigenvalues have discord signs when, in the latter, one replaces $\bar{\lambda}_3= -1$.

The latter would describe a  fixed-line of the RG flow at which the theory exhibits an $\mathcal{N}=1$ on-shell supersymmetry $-$ the $U(1)$ gauge field and the neutral fermion $\xi$  are combined into $\mathcal{N}=1$ gauge supermultiplet which couples to two chiral supermultiplets $(\phi_1, \psi_1)$ and $(\phi_2, \psi_2)$ via gauge interactions. However, it is neither UV attractive nor IR attractive due to the above signs discordance.

The case with concord signs represents an emergent global $U(2)$ symmetry.\\

\section{Emergent  gauge symmetry}

All the known fundamental interactions rely on the principle of local gauge invariance. This is the only known consistent and manifestly Lorentz-invariant description of quantum fields carrying spin $\geq 1$. Yet, one may entertain the possibility that gauge theories are also emergent. This may be particularly true for gravity, for which the usual notions of locality and smooth spacetime manifolds fail at short scales.

\subsection{Emergent $U(1)$ gauge theory }

As a simple example of emergent local gauge theory consider a vector field $A_{\mu}$ coupled to a current $j_{\mu}$. The theory is described by the following Lagrangian:
\begin{equation}
\mathcal{L}=\frac{1}{2}(\partial_{\mu}A^{\mu})^2-\frac{g}{2}(\partial_{\mu}A_{\nu})^2-g'j_{\mu}\,A^{\mu}~,
\label{1}
\end{equation}
 where $g(\mu)$ and $g'(\mu)$ are dimensionless running parameters defined at the renormalization scale $\mu$. It is convenient to decompose the 4-vector potential as:
 \begin{eqnarray}
 A_{\mu}=a_{\mu}+\frac{1}{\Lambda}\partial_{\mu}\phi~,
 \label{2}
 \end{eqnarray}
where $a_{\mu}$ is a divergenceless 4-vector field, $\partial_{\mu}a^{\mu}=0$, $\phi$ is a scalar field and $\Lambda$ is an arbitrary parameter of mass dimension 1, which is introduced to measure $\phi$ in units of $\Lambda$. After rescaling the fields, $a_{\mu}\to \sqrt{g}\,a_{\mu}$ and $\phi\to \sqrt{g-1}\,\phi$, the Lagrangian (\ref{1}) takes the form:
\begin{equation}
\mathcal{L}=-\frac{1}{2}(\partial_{\mu}a_{\nu})^2-\frac{g'}{\sqrt{g}}\,j_{\mu}\,a^{\mu} -\frac{1}{2\Lambda^2}\phi\,\Box^2\,\phi+\frac{g'}{\Lambda\,\sqrt{g-1}}\phi\,\partial_{\mu}j^{\mu}~.
\label{3}
\end{equation}
The first two terms alone describe the usual quantum theory of a  massless Abelian gauge field in the Lorenz gauge coupled to the current with the strength $g'/\sqrt{g}$. The last two terms describe the scalar field with a fourth-order derivative kinetic term, which thus carries two remaining degrees of freedom of the generic (non-gauge) 4-vector field. These degrees of freedom are removed if $g=1$, the case where the theory becomes manifestly gauge invariant.

To show that $g=1$ is a fixed point of the theory, we first note that any diagram, with $\phi$ in the internal legs, is finite, due to the $~1/k^4$ behavior of the $\phi$-propagator. In particular, $\phi-j_{\mu}$ coupling receives only finite corrections, and hence is independent of the renormalization scale, i.e.:
\begin{equation}
\frac{g'(\mu)}{\sqrt{g(\mu)-1}}={\rm const.}
\label{4}
\end{equation}
On the other hand, the $a_{\mu}j^{\mu}$ coupling is known to have the trivial fixed point at IR, i.e.:
\begin{equation}
\frac{g'(\mu)}{\sqrt{g(\mu)}}\stackrel{\mu\to 0}{\longrightarrow}0.
\label{5}
\end{equation}
The equations (\ref{4}) and (\ref{5}) then imply:
\begin{equation}
g(\mu)\stackrel{\mu\to 0}{\longrightarrow}1~,
\label{6}
\end{equation}
that is, the theory asymptotically becomes gauge invariant in the infrared. In the gauge invariant limit (\ref{6}), longitudinal and ghost degrees of freedom become non-dynamical, and $\phi$ in the last term of Eq. (\ref{3}) serves as a Lagrange multiplier field, which enforces the 4-current conservation, $\partial_{\mu}j^{\mu}=0$.

\subsection{Pauli-Fierz flow in generic massive spin-2 theory}

The above discussion can be applied also to emergent non-linearly realized gauge symmetries. As an interesting example consider the diffeomorphism invariant linearized theory of the spin-2 field with the addition of a generic mass term:
\begin{eqnarray}
\mathcal{L}&=&-\frac{1}{2}\,m^2\left(h_{\mu\nu}h^{\mu\nu}-\alpha\, h^2\right)+\beta\, h_{\mu\nu}T^{\mu\nu}+ \nonumber \\
&&\text{gauge invariant terms}~,
\label{7}
\end{eqnarray}
where $h=h^{\mu}_{\,\mu}$. For $\alpha=1$ this theory exhibits non-linearly realized gauge invariance (linearized diffeomorphisms), owing to which there are five propagating degrees of freedom of massive spin-2 field \cite{Fierz:1939ix}. For $\alpha\neq 1$ the theory is not gauge invariant and an additional scalar ghost degree of freedom appears in the spectrum. To see that the asymmetric theory indeed flows towards the $\alpha=1$ Pauli-Fierz theory, let us split the tensor field into the gauge invariant transverse and traceless tensor $h_{\mu\nu}^{\rm TT}$ and the vector field $\xi_{\mu}$:
\begin{equation}
h_{\mu\nu} =h_{\mu\nu}^{\rm TT}+\frac{1}{\Lambda}
\partial_{( \mu}\xi_{\nu)}~.
\label{8}
\end{equation}
The Lagrangian (\ref{7}) then becomes:
\begin{eqnarray}
 \mathcal{L}&=&-\frac{m^2}{\Lambda^2}\left(\partial_{\mu}\xi_{\nu}\right)^2+ \frac\alpha2\,
 \frac{m^2}{\Lambda^2}\left(\partial_{\mu}\xi^{\mu}\right)^2
 -\frac{\beta}{\Lambda}\,\xi_{\nu}\,\partial_{\mu}T^{\mu\nu}
\nonumber \\
 && + \text{ gauge invariant terms}~.
 \label{9}
\end{eqnarray}
The part of this Lagrangian involving the vector field $\xi_{\mu}$ is identical to the one in Eq. (\ref{1}). Hence,  applying the previous analysis, we conclude that the linearized massive spin-2 theory flows towards the ghost-free Pauli-Fierz theory in IR.

\section{Outlook}
The key result of this work is the proposition according to which emergent symmetries are directly related to RG fixed hypersurfaces in the parameter space of {\it a priori} asymmetric theories. We have illustrated the proposition with many simple models with emergent global symmetry, emergent supersymmetry, and emergent gauge symmetry. We would like to believe that these toy models can be expanded to realistic physical theories that address some important phenomenological problems.

The radiative stability of some measured parameters in particle physics and cosmology, most notably of the electroweak scale (i.e. the Higgs mass) and the cosmological constant, is believed to require certain (albeit approximate) symmetries at respective scales. The prime candidates for such symmetries are supersymmetry and scale invariance. It would be interesting to contemplate whether the relevant symmetries that ensure the radiative stability of these parameters are emergent rather than a fundamental feature of the theory.

Emergent symmetries could potentially provide an important technical tool for addressing physical problems beyond the perturbative level. For example, several aspects of the dynamics in the strongly coupled regime can be understood within supersymmetric theories, while the phenomenologically relevant theory, Quantum Chromodynamics, does not expose such symmetry at the fundamental level. Would it be possible to understand the QCD confinement via emergent symmetries in the strongly coupled regime?

Finally, one may think of the gauge symmetries themselves, and most notably the diffeomorphism invariance of gravity, as emergent symmetries. Needless to say, that progress in any of these directions may result in a paradigm-changing discovery.

\section*{Acknowledgements}
The authors acknowledge the partial support from the INFN and the Simons Foundation (grant 341344, LA)
during their visit at the Galileo Galilei Institute for Theoretical Physics, where this work was initiated.
The work was supported in part by the program PRIN 2017 of
Ministero dell'Istruzione, Universit\`a e della Ricerca (MIUR), Grant 2017X7X85K
``The dark universe: synergic multimessenger approach", 
and in part by the Shota Rustaveli National Science Foundation (SRNSF) of Georgia,
Grant DI-18-335/New Theoretical Models for Dark Matter Exploration.

\end{document}